%% file: main.tex
\documentclass[11pt,a4paper]{article}
\usepackage{braket}
\usepackage[blocks]{authblk}
\usepackage{bm}
\usepackage{graphicx}
\usepackage{here}
\usepackage{amsmath}
\usepackage{amssymb}
\usepackage{arydshln}
\usepackage[
dvipdfm,truedimen,%
top=30truemm,bottom=30truemm,%
left=25truemm,right=25truemm]{geometry}

\begin{document}
\title{\bf{Connection of four-dimensional Langevin model and Hauser-Feshbach theory to describe
statistical decay of fission fragments}} 
\author[1]{Kazuki Fujio}
\author[2]{Shin Okumura}
\author[3]{Chikako Ishizuka}
\author[3,4]{Satoshi Chiba}
\author[3]{Tatsuya Katabuchi}

\affil[1]{Nuclear Engineering Course, Department of Transdisciplinary Science and Technology, School of Environment and Society, Tokyo Institute of Technology, 2-12-1 Ookayama, Meguro, 152-8550, Tokyo, Japan,}
\affil[2]{NAPC-Nuclear Data Section, International Atomic Energy Agency, Vienna International Centre, 1400, Vienna, Austria,}
\affil[3]{Institute of Innovative Research, Tokyo Institute of Technology, 2-12-1 Ookayama, Meguro, 152-8550, Tokyo, Japan,}
\affil[4]{NAT Research Center, NAT, 3129-45 Hirahara, Muramatsu, Tokai, Naka, 319-1112, Ibaraki, Japan}

\maketitle

\begin{abstract}
We developed a method superposing two different fission modes calculated in a four-dimensional Langevin model to obtain more accurate fission fragment yield and total kinetic energy (TKE).
The two fission modes correspond to the standard I and standard II modes reported by Brosa et al., and parameters in the Langevin model and the superposing ratio were determined to reproduce the fission fragment yield of $^{240}$Pu of spontaneous fission.
We also investigated the fission fragment yields and the TKEs of other Pu isotopes by using the same Langevin parameters and different superposing ratios, such as spontaneous fission of $^{238,242}$Pu and neutron-induced fission of $^{239}$Pu.
The prompt fission observables, such as the neutron multiplicity, the prompt fission neutron spectrum, and the independent fission product yield were calculated in the Hauser-Feshbach statistical decay model implemented in a nuclear reaction code TALYS with $^{239}$Pu(n,f) in the incident energies ranging from thermal energy up to 5 MeV.
The calculated fission observables qualitatively reproduce several known trends while calculated results have quantitative discrepancies between reported data.
Although more improvements are needed for the most important nuclides, it turned out that our approach has the capability to provide prompt fission observables for difficult-to-measure nuclides.
\end{abstract} 

\bigskip

\section{Introduction}\label{sec1}
\input{introduction}

\section{Models and calculation method}\label{sec2}
\input{method}

\section{Results and discussion}\label{sec3}
\input{result}

\section{Conclusions}\label{4}
\input{conclusion}

\section*{Acknowledgement}
\input{Acknowledgements}

\bibliography{references}
\bibliographystyle{unsrt}

\end{document}

%% file: introduction.tex
Accurate evaluation of fission observables is essential for 
accurate design of nuclear reactors and for nuclear transmutation technology such as accelerator-driven systems~\cite{Tsujimoto2007}.
In particular, the mass distribution of fission products (after prompt neutron emission) and neutrons emitted from fission fragments (before prompt neutron emission) have been studied in detail both experimentally and theoretically.
The independent fission product yield, namely, the mass distribution of fission products after prompt particle emissions, determines the total amount of radioactive waste in spent nuclear fuel and affects disposal scenarios of radioactive waste~\cite{Mukaiyama2001}.
The prompt neutron multiplicity, i.e., the average number of emitted prompt neutrons, is important for evaluating the criticality of nuclear reactors, and the prompt fission neutron spectrum (PFNS) is also essential for criticality and analysis of burn-up characteristics~\cite{Chiba2017}.
For each fission observable, there are several theoretical efforts relying on empirical models~\cite{Wahl1988,Wahl2002,Katakura2003} and models based on nuclear physics~\cite{Madland1982}, respectively.
Currently, approaches using Monte Carlo samplings such as CGMF~\cite{Lemaire2005,Talou2021}, FIFRELIN~\cite{Litaize2010}, FREYA~\cite{Randrup2009,Vogt2009}, and GEF~\cite{Schmidt2016}, or deterministic approaches such as DSE~\cite{Tudora2018}, HF$^3$D~\cite{Okumura2018}, PbP~\cite{Tudora2006}, and TALYS~\cite{Koning2012} have been developed to evaluate fission observables in a consistent manner since they are mutually correlated.

The fission reaction is a dynamical process caused by nuclear deformation and can be described by several different steps.
In the case of neutron-induced fission, an excited compound nucleus undergoes shape deformation by the motion of its internal nucleons.
After surmounting multiple fission barriers, the compound nucleus forms a narrow neck.
The compound nucleus is divided into two highly excited fission fragments by the scission of the neck, and more than a thousand types of fragments are produced.
After scission, the fission fragments proceed to the prompt decay process to release their excitation energy.
In the prompt decay process, the fragments are de-excited to their ground states or the isomeric states by emitting prompt neutrons and $\gamma$ rays.
Then $\beta^-$ decay takes place in neutron-rich independent fission products towards the final stable or long-lived cumulative fission products.

As described above, nuclear fission starts with nuclear deformation and involves several complex physical processes which include the prompt decay and $\beta^-$ decay processes.
Since the fission process consists of several different physical mechanisms, it is necessary to combine multiple theories to describe the entire process.
For the process up to scission, approaches describing the potential energy of a compound nucleus have been developed by using microscopic models~\cite{Sadhukhan2016,Tanimura2017,Lemaitre2019,Bulgac2019,Bulgac2020,Zhao2023} and macro-microscopic models~\cite{Asano2004,Randrup2011,Aritomo2014,Pasca2016,Sierk2017,Ishizuka2017,Jaffke2018,Carjan2019,Mumpower2020}.
The Langevin approach is one of the methods that can describe part of the fission process by using the macro-microscopic potential~\cite{Asano2004,Aritomo2014,Sierk2017,Ishizuka2017}.
The Langevin equation is a stochastic differential equation of motion that describes the Brownian motion of collective variables in a heat bath formed by nucleons, taking into account the effect of the random force acting on macroscopic coordinates.
In nuclear physics, the Langevin model simulates the fission process from nuclear deformation after forming a compound nucleus up to scission.
In our four-dimensional Langevin model, we have succeeded in understanding mode transitions of the fission fragment yield and the total kinetic energy (TKE) distribution over a wide mass range from actinide to superheavy nuclei~\cite{Ishizuka2017,Usang2019,Ishizuka2020}.
However, we need more precise data for the width and peak positions of the mass distribution for nuclear applications.

In this paper, we developed a new method to calculate the fission fragment yield and the TKE more accurately by superposing two different fission modes. 
We also show that this approach has the capability to evaluate the fission fragment data and the prompt fission observables of neutron-induced fission without dependence on the mass and charge number.
Moreover, we performed the Hauser-Feshbach statistical decay calculation and evaluated the prompt fission observables of neutron-induced fission of $^{239}$Pu ($^{239}$Pu(n,f)) by using a nuclear reaction code TALYS with the Langevin fission fragment data (the fission fragment yield and TKE).
This comprehensive approach enables the consistent calculation of the prompt fission observables based on nuclear physics.

In Section 2, we introduce the four-dimensional Langevin model, our approach by superposing two different fission modes, and how we prepare the input for the Hauser-Feshbach statistical decay calculation.
In Section 3, we show the fission fragment yields and TKEs obtained from the four-dimensional Langevin model and the prompt fission observables obtained from TALYS using the Langevin fission fragment data.
Conclusions are given in Section 4.

%% file: method.tex
\subsection{Four-dimensional Langevin model}
\label{sec:Lang}
We employed the four-dimensional Langevin model~\cite{Ishizuka2017,Usang2019,Ishizuka2020} for simulating the deformation of a compound nucleus up to scission.
The time evolution of nuclear deformation is solved in the Langevin equation on the corresponding potential energy with transport coefficients:
\begin{eqnarray}
\label{eq:4D-Langevin}
\frac{dq_{\mu}}{dt}&=&\left(m^{-1}\right)_{\mu\nu}p_{\nu}, \nonumber\\
\frac{dp_{\mu}}{dt}&=&-\frac{\partial F(q,T)}{\partial q_{\mu}}-\frac{1}{2}\frac{\partial\left(m^{-1}\right)_{\nu\sigma}}{\partial q_{\mu}}p_{\nu}p_{\sigma}-\gamma_{\mu\nu}\left(m^{-1}\right)_{\nu\sigma}p_{\sigma}+\sqrt{T_\mu^{\rm eff}} g_{\mu\nu}R_{\nu}(t),
\end{eqnarray}
where $\{q_{\mu} : \mu=1 \cdots 4\}$ is a set of collective variables of nuclear shape, and $\{p_{\mu}\}$ is the corresponding conjugate momenta.  We do not take the sum on $\mu$ in the last term of the second equation.  The symbol $R_\nu$ represents a stochastic force having a white-noise nature.

\begin{figure}[H]
\centering
\includegraphics[width=0.6\textwidth]{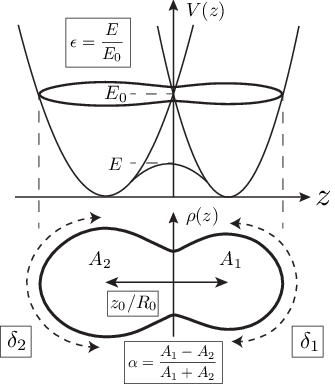}
\caption{The potential of the two-center shell model and the nuclear shape.}
\label{fig:TCSM}
\end{figure}

The time-dependent collective variables $\{q_{\mu}\}=\{{z_0}/R_0, \delta_1, \delta_2, \alpha\}$ and a neck parameter $\epsilon$ are represented in the two-center shell model (TCSM)~\cite{Maruhn1972}.
Figure~\ref{fig:TCSM} shows the potential of the TCSM and the nuclear shape as a function of nuclear elongation.
$z_{0}/R_{0}$ represents nuclear elongation normalized by the radius of the compound nucleus $R_{0}=1.2A^{1/3}$, where $A$ is the mass number of the compound nucleus.
$\delta_{i} (i=1,2)$ corresponds to the deformation of the outer tips of each right and left fragment.
$\alpha$ denotes the mass asymmetry calculated in the difference in mass numbers of each fragment.
The neck configuration is described by a parameter $\epsilon$ using the ratio of the intercept of two harmonic oscillators of the TCSM and that of a connecting function $(\epsilon=E/E_{0})$, where $E_0$ is the actual barrier height.
$\epsilon$ governs the peak positions of the fission fragment yield and overall TKE.
Larger $\epsilon$ gives fission fragments a compact shape, and the peak position shifts to the lighter side in heavy fragments and higher TKE.

$F(q,T)$ in Equation~(\ref{eq:4D-Langevin}) is the temperature-dependent free energy and is employed instead of the potential energy to accurately calculate the temperature dependence of the shell correction. 
$F(q,T)$ is calculated as $F=V-TS$, where $V$ is the nuclear potential energy, $T$ is the nuclear temperature, and $S$ is the entropy.  The shell and pairing corrections to the free energy were determined in our 
previous work by a temperature-dependent calculation~\cite{Ivanyuk2018}.
The single-particle energy is calculated in the finite-depth two-center Woods-Saxon potential with parameters from Ref.~\cite{Pashkevich1971}.
The nuclear deformation is introduced into the wave function using the shape of the TCSM expanded by Cassini ovaloids.
Regarding transport coefficients, we calculated a collective inertia tensor $m_{\mu\nu}$ under Werner-Wheeler approximation~\cite{Kelson1964,Davies1976}.
For the friction tensor $\gamma_{\mu\nu}$, we applied wall-and-window formula with a commonly accepted reduction factor $k_{s}=0.27$~\cite{Blocki1978,Sierk1980,Adeev2005}.
As for the random force, the strength of the random force $g_{\mu\nu}$ connects to the friction tensor through the fluctuation-dissipation theorem at 1 MeV, while the effective temperature $T^{\rm eff}_{\mu}$ is defined as below:
\begin{eqnarray}
g_{\mu\sigma}g_{\sigma\nu}=
\gamma_{\mu\nu}, 
\hspace{5mm}
T^{\rm eff}_{\mu}=\frac{1}{2}\hbar\omega_{\mu}\coth\frac{\hbar\omega_{\mu}}{2T},
\end{eqnarray}
where $T$ is the intrinsic temperature of the system and governs the strength of the shell effect.
$T$ is connected to the intrinsic excitation energy $E_{int}$ with the level density parameter $a$:
\begin{eqnarray}
E_{int}=E_{x}-\frac{1}{2}\left(m^{-1}\right)_{\mu\nu}p_{\mu}p_{\nu}-F(q,T=0)=aT^2,
\end{eqnarray}
where $E_{x}$ is the input excitation energy and is treated as the sum of the neutron separation energy and the incident energy. 
$T^{\rm eff}_{\mu}$ is introduced not to drop $T$ beyond the zero-point energy for each collective variable, 
i.e. $T^{\rm eff}_{\mu}=\hbar\omega_{\mu}/2\hspace{2mm}(T\rightarrow0)$, where $\hbar\omega_{\mu}/2$ corresponds to the zero-point energy of the harmonic oscillator.
Larger $\hbar\omega_{4}$ gives wider width of the fission fragment yield and higher ${\rm \braket{TKE}}(A)$ around $A=120-130$.
For the value of parameters $\epsilon$ and $\hbar\omega_{\mu}$ in this research, refer to Section~\ref{sec:Brosa}

In a recent microscopic approach, it is concluded that the collective motion is overdamped, so the inertia is irrelevant~\cite{Bulgac2019,Bulgac2020}.  
In the Langevin approach, however, we describe the collective motion in a different context.  
We start the calculation from a point close to the ground state, and the fluctuations lead the trajectories to pass over the saddle point so that we can obtain ``distributions" of physical quantities such as mass and TKE distributions.  
We need the terms containing the inertia tensor so that our approach to be self-consistent to describe all these processes consistently.  
We hope that the relation between these two different views on the collective motion should be elucidated well in the future.  


\subsection{Superposition of two different fission modes}
\label{sec:Brosa}
In our previous calculations, the width and peak positions of the mass distribution were not well reproduced because the model restricts to one neck parameter $\epsilon$~\cite{Okumura2020}.
The neck parameter $\epsilon$ is a parameter in the four-dimensional Langevin model due to the limitations of calculation time and resources.
However, in the original concept of the TCSM, $\epsilon$ is also supposed to be one of the dynamic variables, and a distribution of $\epsilon$ should exist at the scission point.
From experimental findings, it is known that the fission yields obtained from the fission in actinide nuclides are the shape of a superposition of mass distributions with peaks around $A=134$ and 144.
These peak positions are governed by the shell effects, and the resulting nuclear shapes of the fragments differ accordingly.
To incorporate different fission modes, we superposed two Langevin calculations. 
This approach was inspired by the Brosa model~\cite{Brosa1990} which represents the fission yield as the sum of several fission modes, and this approach enabled us to express the variations in $\epsilon$ in the fission fragment yield and TKE.

Brosa et al. suggested a model on different fission paths: the superlong (SL), supershort (SS), standard I (ST1), and standard II (ST2) modes~\cite{Brosa1990}.
The SL and SS modes represent symmetric fission components and are distinguished by the nuclear shape.
The SL mode gives smaller TKE due to the large deformation, whereas the SS mode produces larger TKE owing to the compact shape.
The ST1 and ST2 modes are asymmetric fission components and are classified by the peak positions, and the shell effect determines their peak positions.
For heavy fragments, the ST1 mode is located near $A=134$ due to the doubly magic shell, and its nuclear shape shows a spherical configuration in the ground state.
The peak position of ST2 mode is in the vicinity of $A=144$ due to the deformed shell, and the shape exhibits deformed.
By using this classification, we fitted the experimental fission fragment yields of $^{240}$Pu(sf) by the Gaussian functions for each ST1 and ST2 mode and performed Langevin calculations to reproduce the Gaussian functions.
The symmetric components are already included in the Langevin calculations and tend to overestimate the fission fragment yield compared to the experimental results in the symmetric region of the ST2 mode.
We did not superpose the SL and SS modes separately and applied a damping function to the symmetric components of the ST2 mode at thermal energy.

As indicated in Section~\ref{sec:Lang}, the peak positions of the fission fragment yield and overall TKE are highly influenced by the parameter $\epsilon$, and the width of the yield and TKE ranging of $A=120-130$ are notably sensitive to the parameter $\hbar\omega_{4}$.
We developed a phenomenological approach adjusting $\hbar\omega_\mu$ and $\epsilon$ to describe the fission fragment yield and average TKE systematically over a wide mass range.
Currently, few fissile isotopes exist in experimental and evaluated data for the fission observables before and after prompt decay.
The fission fragment yields and average TKE distributions are reported for $^{238,240,242}$Pu of spontaneous fission ($^{238,240,242}$Pu(sf)) and $^{239}$Pu of neutron-induced fission ($^{239}$Pu(n,f)) from Ref~\cite{Schillebeeckx1992}.
For this reason, two Langevin parameters are adjusted to reproduce each of the ST1 and ST2 modes for experimental fission fragment yield of $^{240}$Pu(sf).
We extracted the systematics of $\epsilon$ and $\hbar\omega_{\mu}$ and applied the same approach to $^{239}$Pu(n,f).

The parameters are determined to be $\epsilon=0.65$ and $\hbar\omega_{\mu}=(2,2,2,1)$ for the ST1 mode and to be $\epsilon=0.25$ and $\hbar\omega_{\mu}=(2,2,2,2.7)$ for the ST2 mode.
The neck parameter $\epsilon$ differs between the ST1 and ST2 modes because $\epsilon$ reflects the characteristics of nuclear shape, and we confirmed the ST1 mode as a spherical shape and the ST2 mode as a deformed shape at peak positions in heavy fragments.
The characteristics of the width for the mass distribution appear in $\hbar\omega_4$, and that tends to be smaller for the ST1 mode than for the ST2 mode in the actinide nuclides.
We summed up the ST1 and ST2 modes using a superposing ratio $\zeta$ for the fission fragment yield and TKE: 
\begin{eqnarray}
Y_{\rm ff}(A,{\rm TKE})=\zeta Y_{\rm ST1}(A,{\rm TKE})+(1-\zeta)Y_{\rm ST2}(A,{\rm TKE}),
\end{eqnarray}
where $Y_{\rm ST1}(A,{\rm TKE})$ and $Y_{\rm ST2}(A,{\rm TKE})$ are fission fragment yields of ST1 and ST2 modes.
The same $\epsilon$ and $\hbar\omega_{\mu}$ sets are used for $^{238,240,242}$Pu(sf) and $^{239}$Pu(n,f), and we only adjusted $\zeta$ by using a least squares method to reproduce the peak positions of fission fragment yields.

\subsection{Connection to the Hauser-Feshbach statistical decay calculation}
We calculated the prompt decay process with the Hauser-Feshbach theory implemented in TALYS (version:1.96)~\cite{Koning2012} using the fission fragment yields $Y_{\rm ff}(A)$ and average TKE ($\braket{\rm TKE}(A)$) from our Langevin model.
The necessary information for the prompt decay calculation is not only the $Y_{\rm ff}(A)$ and $\braket{\rm TKE}(A)$ but also the charge distribution for each fission fragment, the excitation energy distribution, and the spin-parity distribution for each charge and mass number.  
For the charge distribution, we employed Wahl's $Z_{p}$ model~\cite{Wahl1988,Wahl2002} for each fragment obtained from the Langevin calculations.
The average total excitation energy ($\braket{\rm TXE}$) and its dispersion for each fragment pair are calculated with the Q-value, $\braket{\rm TKE}$, and its dispersion from the Langevin model.
Then, the average excitation energy is distributed into each fission fragment by an anisothermal parameter $R_{T}$~\cite{Ohsawa1999,Ohsawa2000,Okumura2018,Okumura2022}.
$R_T$ is defined as the ratio of effective temperatures $T_{l,h}$ of light and heavy fission fragments:
\begin{eqnarray}
    R_T=\frac{T_l}{T_h}=\sqrt{\frac{U_la_h(U_h)}{U_ha_l(U_l)}},
\end{eqnarray}
where $a(U_{l,h})$ are the level density parameters, and $U_{l,h}$ are the excitation energies corrected by the shell effect.
The energy dependence of $R_T$ is introduced as in previous research~\cite{Okumura2022}:
\begin{align}
R_T=\left\{
\begin{array}{l}
R_{T_{0}}+E_{n}R_{T_{1}},\hspace{3mm} R_{T_{0}}+E_{n}R_{T_{1}}\geq1, \\
1,\hspace{25mm} {\rm otherwise},
\end{array}
\right.
\end{align}
where $R_{T_{0}}$ and $R_{T_{1}}$ are model parameters, and we used $R_{T_{0}}=1.30$ and $R_{T_{1}}=-0.507$ for the $^{239}$Pu(n,f) reaction.

The excitation energy and spin-parity distributions are calculated in TALYS.
TALYS generates a Gaussian distribution of the excitation energy for each fission fragment with the average excitation energy and its dispersion.
The energy and its dispersion are prepared with the Langevin results and the energy-dependent $R_T$ value.
The spin-parity distribution is provided in the form of a Rayleigh distribution.
The parameters in the spin-parity distribution are selected to reproduce the neutron multiplicity at thermal incident energy.
For further details, please refer to the relevant reference~\cite{Fujio2023}.

%% file: result.tex
\subsection{Fission fragment yield and average TKE obtained from four-dimensional Langevin model}
\begin{figure}[H]
\centering
\includegraphics[width=0.9\textwidth]{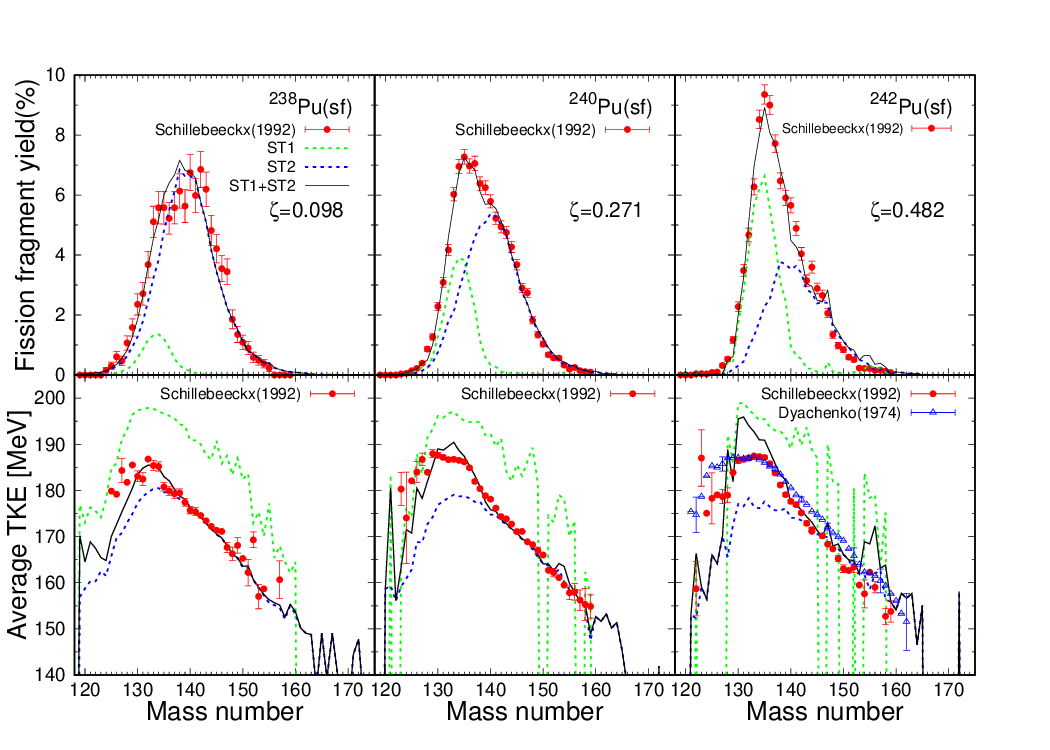}
\caption{(Upper) The calculated fission fragment yield $Y_{\rm ff}(A)$ and (lower) the average TKE ($\braket{\rm TKE}(A)$) for $^{238, 240, 242}$Pu(sf). The green dotted lines are calculated ST1 modes ($\epsilon=0.65, \hbar\omega_{\mu}=(2,2,2,1)$), the blue dotted lines are calculated ST2 modes $(\epsilon=0.25, \hbar\omega_{\mu}=(2,2,2,2.7))$, and the black lines are the superposing result of calculated ST1 and ST2 modes.}
\label{fig:input}
\end{figure}

We calculated the fission fragment yields $Y_{\rm ff}(A)$ and the average TKEs ($\braket{\rm TKE}(A)$) of both ST1 and ST2 modes with the Langevin model to evaluate fission observables with the prompt decay calculation.
The upper part of Figure \ref{fig:input} shows $Y_{\rm ff}(A)$, and the lower part of Figure~\ref{fig:input} represents $\braket{\rm TKE}(A)$ for $^{238,240,242}$Pu(sf).
By adjusting the superposing ratio, the peak position of the $Y_{\rm ff}(A)$ is successfully reproduced for $^{238, 240, 242}$Pu(sf). 
The calculated $\braket{\rm TKE}(A)$ is also in agreement with experimental data. 
This agreement is attributed to the combination of both ST1 and ST2 modes.
The ST1 mode has a larger $\braket{\rm TKE}(A)$ than the ST2 mode because the shape of the fission fragment is more spherical in the ST1 mode.
Table \ref{tab:TKE} shows the calculated and experimental $\braket{\rm TKE}(A)$ for Pu isotopes.
The deviation between our results and the experimental data is approximately 2\%, which indicates our approach can provide relatively accurate $\braket{\rm TKE}$ simultaneously with $Y_{\rm ff}(A)$.

\begin{table}[H]
\centering
\caption{The calculated and experimental average TKE ($\braket{\rm TKE}$) for Pu isotopes at thermal energy (in MeV unit).}
\begin{tabular}{lllll}\hline
&& Standard I & Standard II & Average TKE \\
&The neck parameter $\epsilon$& $0.65$ & $0.25$ & \\ 
&The zero-point energy $\hbar\omega_{\mu}$& $(2,2,2,1)$ & $(2,2,2,2.7)$ & \\  \hline
 $^{238}$Pu(sf) & Langevin calc. & 195.87 & 174.36 & 176.47   \\ 
   & Dematt\`{e} et al.\cite{Dematte1997} & & & 176.4$\pm$0.3 \\
   & Schillebeeckx et al.\cite{Schillebeeckx1992} & & & 177.0$\pm$0.3  \\ \hline 
 $^{240}$Pu(sf) & Langevin calc. & 194.79 & 173.44 & 179.23   \\ 
   & Dematt\`{e} et al.\cite{Dematte1997} & & & 178.5$\pm$0.1 \\
   & Schillebeeckx et al.\cite{Schillebeeckx1992} & & & 179.4$\pm$0.1 \\ \hdashline 
  $^{239}$Pu(n$_{\rm th}$,f) & Langevin calc. & 194.78 & 173.65 & 176.88  \\ 
    &Wagemans et al.\cite{Wagemans1988,Wagemans1989} & 192 & 175 &  \\
    &Schillebeeckx et al.\cite{Schillebeeckx1992} &  &  & 177.93$\pm$0.01 \\
    &Surin et al.\cite{Surin1972} &  &  & 177.7$\pm$0.1 \\
    &Tsuchiya et al.\cite{Tsuchiya2000} &  &  & 176.2$\pm$1.4 \\  \hline
 $^{242}$Pu(sf) & Langevin calc. & 194.22 & 172.57 & 183.01   \\ 
   & Dematt\`{e} et al.\cite{Dematte1997} & & & 180.5$\pm$0.1 \\
   & Schillebeeckx et al.\cite{Schillebeeckx1992} & & & 180.7$\pm$0.1 \\ \hline
\end{tabular}
\label{tab:TKE}
\end{table}

\begin{figure}[H]
\centering
\includegraphics[width=0.7\textwidth]{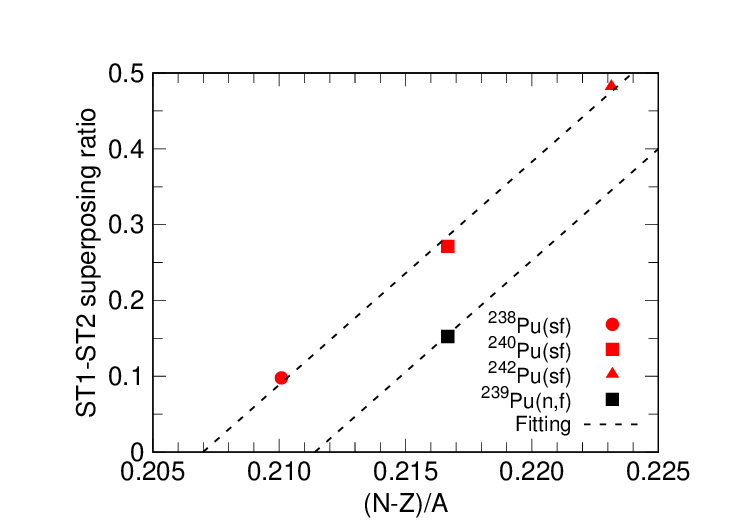}
\caption{The superposing ratio $\zeta$ of the ST1 and ST2 modes as a function of $(N-Z)/A$ for Pu isotopes. The dashed line represents a fitting function calculated from Pu(sf) isotopes.}
\label{fig:ratio}
\end{figure}

Figure~\ref{fig:ratio} represents the superposing ratio $\zeta$ of the ST1 and ST2 modes as a function of $(N-Z)/A$ for Pu isotopes.
We can see that $\zeta$ is in proportion to $(N-Z)/A$ for spontaneous fission.
This result shows the amount of $Y_{\rm ST1}(A)$ increases linearly as a function of $(N-Z)/A$.
Assuming that the slope of $\zeta$ for neutron-induced fission is the same as that for spontaneous fission of Pu isotopes, the superposing approach might have the capability to calculate $Y_{\rm ff}(A)$ and TKE for other nuclides.

\begin{figure}[H]
\centering
\includegraphics[width=0.9\textwidth]{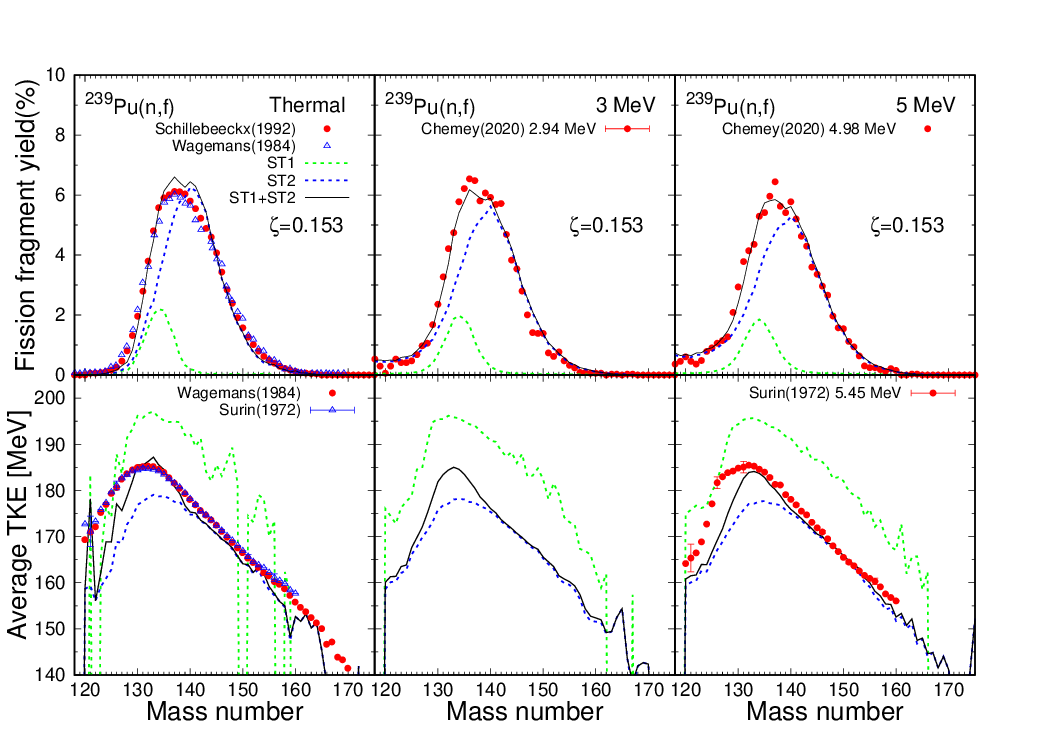}
\caption{(Upper) The calculated fission fragment yield $Y_{\rm ff}(A)$ and (lower) the average TKE ($\braket{\rm TKE}(A)$) for $^{239}$Pu(n,f). The green dotted lines are calculated ST1 modes ($\epsilon=0.65, \hbar\omega_{\mu}=(2,2,2,1)$), the blue dotted lines are calculated ST2 modes $(\epsilon=0.25, \hbar\omega_{\mu}=(2,2,2,2.7))$, and the black lines are the superposing result of the calculated ST1 and ST2 modes.}
\label{fig:input_Pu239}
\end{figure}

We applied the same $\epsilon$ and $\hbar\omega_{\mu}$ to $^{239}$Pu(n,f), and Figure~\ref{fig:input_Pu239} represents $Y_{\rm ff}(A)$ and $\braket{\rm TKE}(A)$ for $^{239}$Pu(n,f) at thermal energy and at 3 and 5 MeV of incident energies.
$\zeta$ is adjusted to reproduce the experimental fission fragment yield of $^{239}$Pu(n$_{\rm th}$,f), and the same $\zeta$ is employed for the other incident energies. 
The increase of $Y_{\rm ff}(A)$ in the symmetric region with increasing the incident neutron energy is well reproduced.
The peak position and the width of each fission fragment yield $Y_{\rm ff}(A)$ are generally in good agreement with the experimental data although the peak position is out by a few mass numbers and is overestimated in the vicinity of $A=137, 140$, especially at thermal incident energy.
Regarding $\braket{\rm TKE}(A)$, it is seen that the calculation results successfully reproduce the experimental data in $A\geq130$ while they underestimated around $A=120-130$.
To improve the accuracy of $\braket{\rm TKE}(A)$, more $Y_{\rm ff}(A)$ should be obtained in the region of $A=120-130$ in the ST1 mode.

\begin{figure}[H]
\centering
\includegraphics[width=0.7\textwidth]{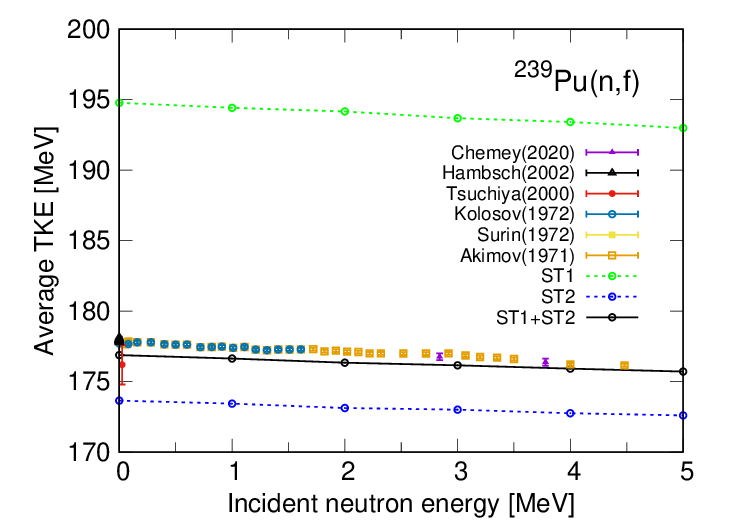}
\caption{The average TKE ($\braket{\rm TKE}(E)$) as a function of incident energy for $^{239}$Pu(n,f). The green dotted lines are calculated ST1 modes, the blue dotted lines are calculated ST2 modes, and the black lines are the superposing result of the calculated ST1 and ST2 modes.}
\label{fig:TKE}
\end{figure}

Figure~\ref{fig:TKE} shows $\braket{\rm TKE}(E)$ as a function of incident neutron energy ranging from thermal up to 5 MeV.
The calculation indicates the underestimation of $\braket{\rm TKE}(E)$ at thermal energy especially, but the decreasing trend is reproduced in $\braket{\rm TKE}(E)$ as the incident energy increases.
One of the reasons for the underestimation in $\braket{\rm TKE}(E)$ is the underestimation of $\braket{\rm TKE}(A)$ around $A=120-130$. 
In terms of the decreasing slopes, there is a discrepancy between the experimental and calculated results.
The previous research reported that $\braket{\rm TKE}$ in Langevin calculation decreases as increasing the incident neutron energy due to the nuclear deformation~\cite{Shimada2021}.
Therefore, the decreasing slopes in Langevin calculations are decided in the balance between the nuclear deformation and the amount of $Y_{\rm ff}(A)$ in the symmetric region.
On the other hand, there are some uncertainties in the experimental TKE because the experimental TKE is obtained after processed by theoretical or phenomenological procedures to correct the prompt decay process.
Moreover, the number of experimental TKE results is limited in actinide nuclei.
We cannot conclude the reasons for the discrepancies of the decreasing slope between the calculated and experimental $\braket{\rm TKE}(E)$.
The investigation of TKE using the Langevin model will be conducted in the future.

\subsection{Prompt fission neutron observables and independent fission product yield calculated in TALYS}
\begin{figure}[H]
\begin{minipage}{0.5\textwidth}
\centering
\includegraphics[width=\textwidth]{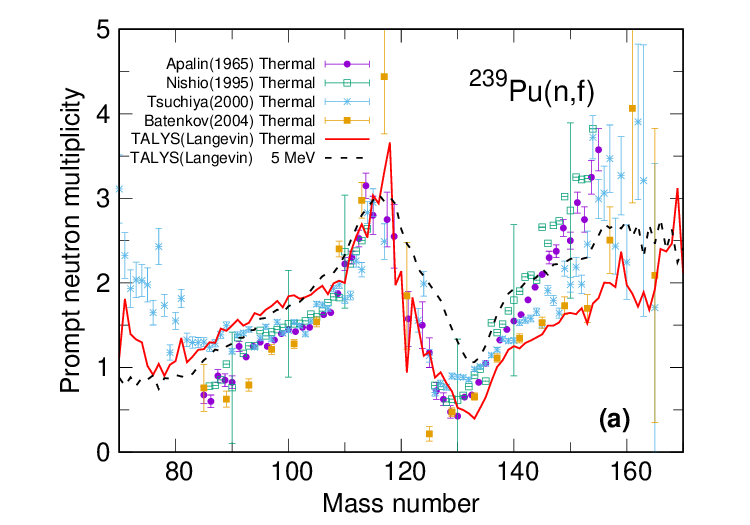}
\end{minipage}
\begin{minipage}{0.5\textwidth}
\centering
\includegraphics[width=\textwidth]{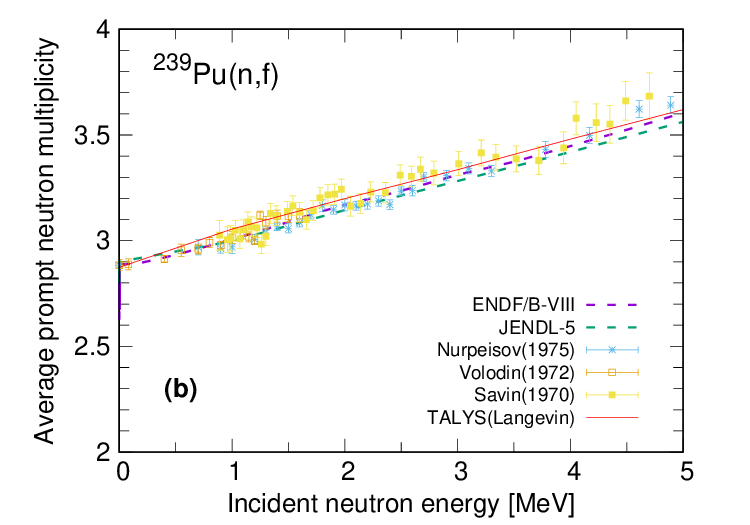}
\end{minipage}
\caption{ (a) The neutron multiplicity $\bar{\nu}_n(A)$ as a function of mass number at thermal and 5 MeV and (b) $\bar{\nu}_n(E)$ as a function of incident energy ranging from thermal up to 5 MeV for $^{239}$Pu(n,f).}
\label{fig:nun}
\end{figure}

By using the Hauser-Feshbach statistical decay calculation implemented in TALYS, we investigated the fission neutron observables and the independent fission product yield.
Figure~\ref{fig:nun}(a) shows the prompt fission neutron multiplicity $\bar{\nu}_n(A)$ for $^{239}$Pu(n,f) as a function of mass number at the incident energies of the thermal energy and 5 MeV.
$\bar{\nu}_n(A)$ reproduces successfully the known tendency while calculated results overestimate (underestimate) the experimental data in the light (heavy) fragments.
At thermal energy, the calculated $\bar{\nu}_{n}(A)$ shows the saw-tooth shape as widely known in experimental results for actinide nuclei.
Compared to $\bar{\nu}_{n}(A)$ at the incident energy of 5 MeV, $\bar{\nu}_{n}(A)$ increases mainly from the heavy fragments owing to the energy-dependent $R_T$ value.
$\bar{\nu}_{n}(A)$ increase from heavy fragments has been reported for several actinide nuclei in Ref.~\cite{Mueller1984,Naqvi1986} for experimental results, and several theoretical codes such as GEF, PbP~\cite{Tudora2015}, FIFRELIN~\cite{Thulliez2019}, and CGMF with time-dependent superfluid local density approximation (TDSLDA) results~\cite{Bulgac2019,Bulgac2020} also support this trend.

Figure~\ref{fig:nun}(b) shows $\bar{\nu}_{n}(E)$ for $^{239}$Pu(n,f) as a function of the incident energy ranging from thermal up to 5 MeV, and the calculated $\bar{\nu}_{n}$ at thermal energy is tabulated with experimental and evaluated data in Table~\ref{tab:nun}.
The slope is different between the range from 0 to 1 MeV and that from 1 to 5 MeV due to the damping function for the symmetric components at thermal energy.
The slope from 0 to 1 MeV is steeper because of increasing $\bar{\nu}_{n}$ from the symmetric fission fragments.
The calculated $\bar{\nu}_{n}(E)$ is in fairly good agreement with experimental and evaluated data ranging from thermal up to 5 MeV even though there is a discrepancy of 1\% in the calculated and experimental TKE.
There are several reasons why calculated $\bar{\nu}_{n}(E)$ reproduces the reported data:
(1) The discrepancy of the TKE is approximately 1 MeV, and it is small compared to the threshold energy of the prompt neutron evaporation, i.e., the neutron separation energy.
(2) $\bar{\nu}_{n}(E)$ varies with not only the excitation energy but also other conditions such as the spin-parity distribution of the fission fragments, the fission fragment yield, and the $R_T$ value.
Various conditions overlapped, and the calculated result is in good agreement with the known data.
(3) In our calculation, the overestimation and underestimation in $\bar{\nu}_{n}(A)$ cancel out each other, consequently, $\bar{\nu}_{n}(E)$ reproduces the reported data.
Thus, we need to take into account several input conditions more carefully.

\begin{table}[H]
\centering
\caption{The calculated neutron multiplicity $\bar{\nu}_{n}$ at thermal energy for $^{239}$Pu(n,f).}
\begin{tabular}{llll}\hline
  & TALYS(Langevin) & ENDF-B/VIII.0 & JENDL-5  \\ \hline
  & 2.874 & 2.870 & 2.870   \\ \hline
\end{tabular}
\label{tab:nun}
\end{table}

\begin{figure}[H]
\centering
\includegraphics[width=0.68\textwidth]{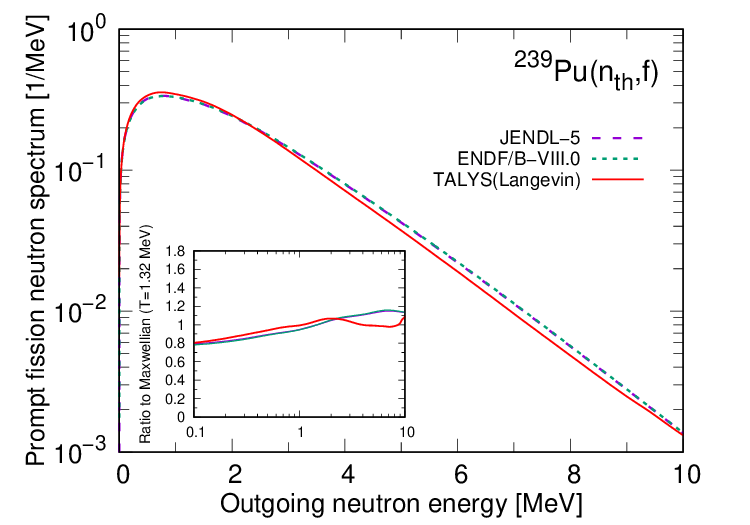}
\caption{The calculated prompt fission neutron spectrum (PFNS) in the laboratory frame for $^{239}$Pu(n$_{\rm th}$,f). The inset figure represents the PFNS as a ratio to a Maxwellian spectrum at $T=1.32$ MeV.}
\label{fig:PFNS}
\end{figure}

Figure~\ref{fig:PFNS} shows the calculated prompt fission neutron spectrum (PFNS) in the laboratory frame, and the inset is that of a ratio to a Maxwellian spectrum.
Although the calculated PFNS underestimates the evaluated data from 3 MeV up to 10 MeV, the calculated one approximately reproduces the shape of the evaluated ones on a logarithmic scale.
The PFNS is influenced by the spin-parity distributions of fission fragments~\cite{Kawano2021,Fujio2023}, and it is known that the fission fragment yield also affects the tail of the PFNS~\cite{Kawano2021}.
It is necessary to determine the fission fragment yield and the spin-parity distribution while keeping the accuracy of other fission observables, such as the neutron multiplicity and the independent fission product yield.
We have not found the solution for it, and the investigation for the PFNS and other fission observables is in progress.

\begin{figure}[H]
\centering
\includegraphics[width=0.68\textwidth]{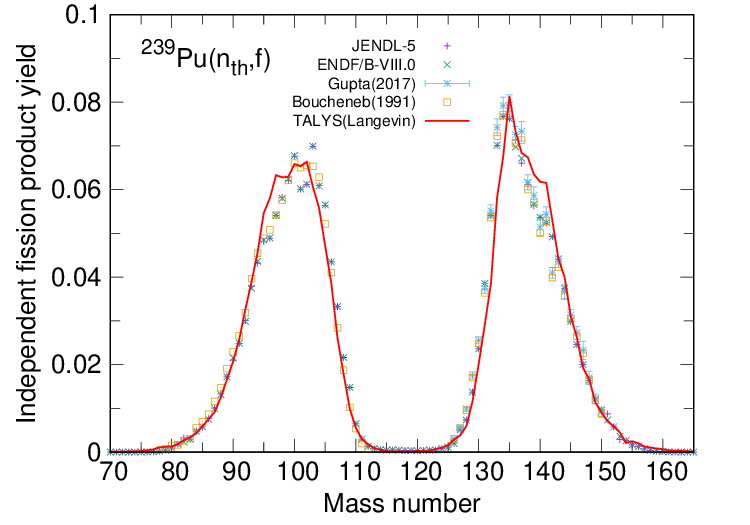}
\caption{Comparison of the calculated independent fission product yield $Y(A)$ with the experimental and evaluated data for $^{239}$Pu(n$_{\rm th}$,f).}
\label{fig:yield}
\end{figure}

Figure~\ref{fig:yield} represents the independent fission product yield $Y(A)$ for $^{239}$Pu(n,f) at thermal energy.
The accuracy of $Y(A)$ has improved from the previous research~\cite{Okumura2020} by employing two Langevin calculations of ST1 and ST2 modes.
While the current approach partially reproduces fine structure in $Y(A)$, the peak positions are slightly out by a few mass numbers compared to the experimental and evaluated data.
The calculated $Y(A)$ is overestimated in the vicinity of $A=97$ for the light fragment and $A=141$ for the heavy fragment.
The overestimations in $Y(A)$ are derived from the overestimations in the fission fragment yield $Y_{\rm ff}(A)$ in the vicinity of $A=137, 140$.
The calculation suggests that an accurate fission fragment yield $Y_{\rm ff}(A)$ is necessary to get a precise independent fission product yield $Y(A)$.

\begin{figure}[H]
\begin{minipage}{0.5\textwidth}
\centering
\includegraphics[width=\textwidth]{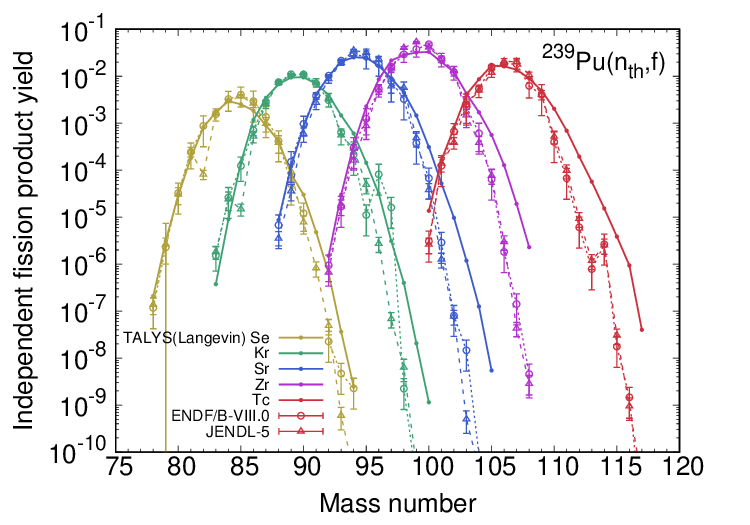}
\end{minipage}
\begin{minipage}{0.5\textwidth}
\centering
\includegraphics[width=\textwidth]{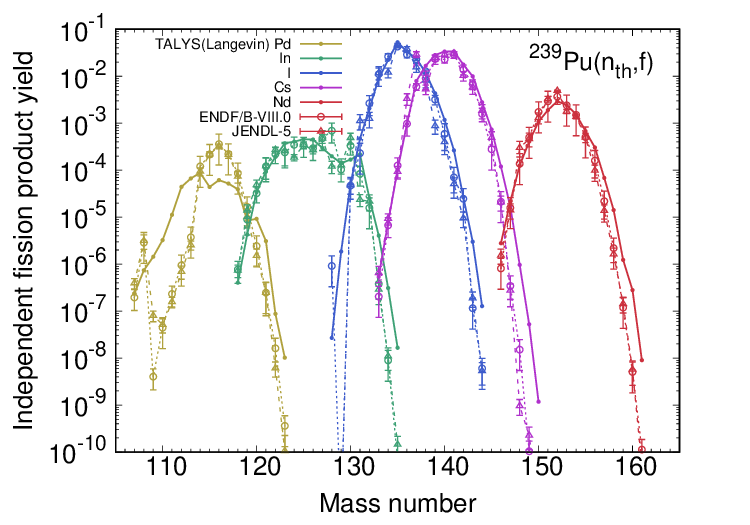}
\end{minipage}
\caption{The calculated independent fission product yield $Y(Z,A)$ for several isotopes of $^{239}$Pu(n$_{\rm th}$,f).}
\label{fig:zyield}
\end{figure}

Figure~\ref{fig:zyield} illustrates the independent fission product yields $Y(Z,A)$ as functions of charge and mass numbers.
Our investigation focused on specific isotopes and revealed a notable discrepancy in Pd isotopes compared to the evaluated data.
This discrepancy arises from the fact that the determination of $Y_{\rm ff}(A)$ relies on a least squares method, and the small amount of $Y_{\rm ff}(A)$ in the symmetric region has not been adjusted to reproduce the known data.
Consequently, these results emphasize the necessity of modifications within the symmetric region to enhance the accuracy of both $Y(Z,A)$ and $Y_{\rm ff}(A)$.
For the other isotopes, the calculated $Y(Z,A)$ exhibits good agreement with the evaluated data on the lighter side. 
However, we can see the overestimations in $Y(Z,A)$ on the heavier side, i.e., the neutron-rich side.
To further investigate this phenomenon, we investigated the charge distribution of $Y(Z,A)$ at specific mass numbers, namely $A=100, 103, 134$, corresponding to the characteristic peaks in the evaluated $Y(A)$.

\begin{figure}[H]
\centering
\includegraphics[width=0.65\textwidth]{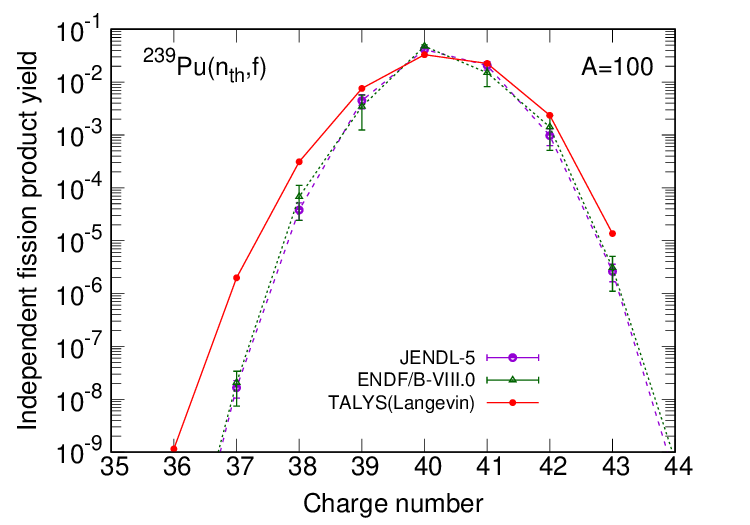}
\includegraphics[width=0.65\textwidth]{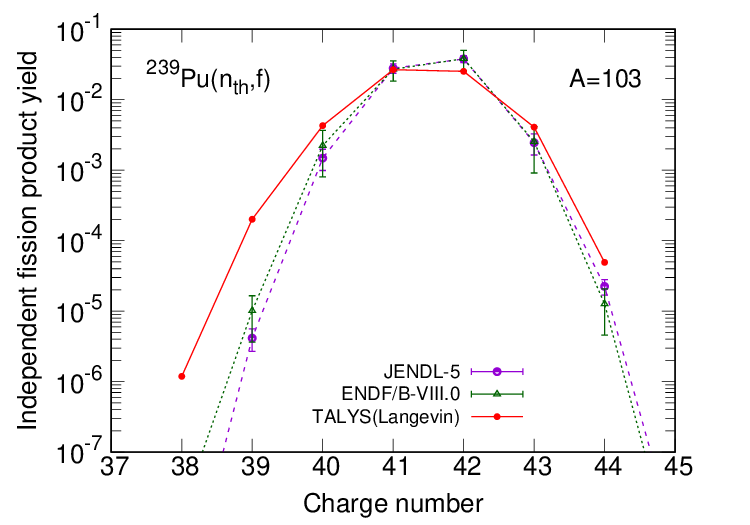}
\includegraphics[width=0.65\textwidth]{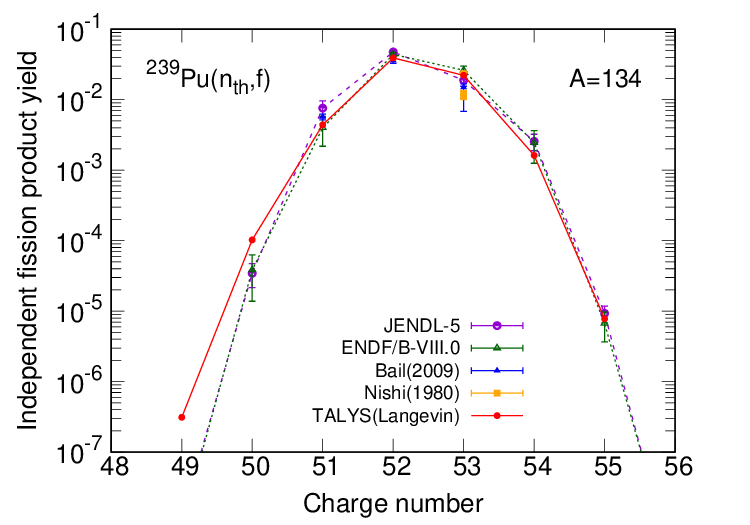}
\caption{Comparison of the calculated independent fission product yield $Y(Z,A)$ with the experimental and evaluated data at $A=100, 103, 134$ for $^{239}$Pu(n$_{\rm th}$,f).}
\label{fig:zyieldA}
\end{figure}

Figure~\ref{fig:zyieldA} shows the calculated $Y(Z,A)$ as a function of charge number $Z$ at $A=100, 103, 134$.
While the calculated $Y(Z,A)$ show a decent agreement with reported data on the heavier charge number side, the results overestimate on the lighter charge number side.
The fission products in the ranges with smaller charge numbers for $Y(Z,A)$ with the same $A$ are more neutron-rich.
We can conclude that $Y(Z,A)$ calculated in our approach tends to overestimate $Y(Z,A)$ on the neutron-rich nuclei by using the original Wahl's $Z_p$ model, especially in $^{239}$Pu(n$_{\rm th}$,f).
The neutron-richness of fission products exerts a significant influence on the neutron emission by $\beta^{-}$ decay, thus an accurate evaluation is necessary.
It clarified the necessity of adjusting the width parameters of the charge distribution to be in good agreement with the evaluated data of $Y(Z,A)$.

%% file: conclusion.tex
We developed a method to calculate accurate fission fragment yields $Y_{\rm ff}(A)$ and total kinetic energies (TKEs) and performed the Hauser-Feshbach statistical decay calculation by using a nuclear reaction code TALYS.
In this method, assuming the fission yield consists of several different fission modes, we superposed two Langevin calculations corresponding to the standard I (ST1) and standard II (ST2) modes.
The current approach successfully describes the arbitrary width and peak position of $Y_{\rm ff}(A)$ by adjusting the neck parameter $\epsilon$ and the zero-point energy $\hbar\omega_{\mu}$, and the accuracy of $Y_{\rm ff}(A)$ and $\braket{\rm TKE}(A)$ was improved as well for $^{238,240,242}$Pu(sf) and $^{239}$Pu(n,f).
The approach also reproduces the incident energy dependence of $Y_{\rm ff}(A)$ in the symmetric region and the decreasing trend in $\braket{\rm TKE}$ for $^{239}$Pu(n,f). 
However, the Langevin calculations still include discrepancies compared to the known data, such as the overestimations on $A=137, 140$ in $Y_{\rm ff}(A)$ and the underestimations on $A=120-130$ in $\braket{\rm TKE}(A)$ and on $\braket{\rm TKE}(E)$.
These discrepancies have large impacts on the prompt fission observables, therefore, we need to improve $Y_{\rm ff}(A)$ and TKE.

Regarding the prompt fission observables of $^{239}$Pu(n,f), it was found that the neutron multiplicity $\bar{\nu}_n(A)$ reproduces the saw-tooth shape and the increasing trend in $\bar{\nu}_n(A)$ by increasing incident energy while $\bar{\nu}_n(A)$ has discrepancies between the experimental results in light and heavy fragments.
The calculated $\bar{\nu}_n(E)$ is also in good agreement with experimental and evaluated data in the range of thermal up to 5 MeV by using the same $R_T$ value as Ref~\cite{Okumura2022}.
The calculated results have discrepancies between the evaluated and experimental data in the prompt fission neutron spectrum (PFNS) and the independent fission product yield $Y(A)$ and $Y(Z,A)$, and these discrepancies are strongly correlated with not only  $Y_{\rm ff}(A)$ and the TKE, but also the spin-parity and charge distributions of fission fragments.
The prompt fission observables are all correlated, it is found that the determination of the input is rather difficult.

Although challenges remain for the most important nuclides, our approach can be useful and accurate enough to provide evaluated data for other nuclides.
We also showed the result that the superposing method might have the capability to calculate $Y_{\rm ff}(A)$ and $\braket{\rm TKE}$ for other nuclides by using the superposing ratio $\zeta$.
The evaluations of $Y_{\rm ff}(A)$ and $\braket{\rm TKE}$ of other neutron-induced systems are in progress.


%% file: Acknowledgements.tex
Authors acknowledge the Grant-in-Aid for Scientific Research(B), MEXT, Japan, and by Japan Society for the Promotion of Science (JSPS) KAKENHI Grant Number 21H01856.

%% file: main.bbl
\begin{thebibliography}{10}

\bibitem{Tsujimoto2007}
K.~Tsujimoto, H.~Oigawa, N.~Ouchi, K.~Kikuchi, Y.~Kurata, M.~Mizumoto, T.~Sasa,
  S.~Saito, K.~Nishihara, M.~Umeno, and Y.~Tazawa.
\newblock Research and development program on accelerator driven subcritical
  system in {JAEA}.
\newblock {\em J Nucl Sci Technol}, 44(3):483--490, 2007.

\bibitem{Mukaiyama2001}
T.~Mukaiyama, T.~Takizuka, M.~Mizumoto, Y.~Ikeda, T.~Ogawa, A.~Hasegawa,
  H.~Takada, and H.~Takano.
\newblock Review of research and development of accelerator-driven system in
  japan for transmutation of long-lived nuclides.
\newblock {\em Prog Nucl Energy}, 38(1):107--134, 2001.
\newblock Accelerator Transmutation of Waste.

\bibitem{Chiba2017}
S.~Chiba, T.~Wakabayashi, Y.~Tachi, N.~Takaki, A.~Terashima, S.~Okumura, and
  T.~Yoshida.
\newblock Method to reduce long-lived fission products by nuclear
  transmutations with fast spectrum reactors.
\newblock {\em Sci Rep}, 7(1):13961, Oct 2017.

\bibitem{Wahl1988}
A.~C. Wahl.
\newblock {Nuclear-charge distribution and delayed-neutron yields for
  thermal-neutron-induced fission of $^{235}$U, $^{233}$U, and $^{239}$Pu and
  for spontaneous fission of $^{252}$Cf}.
\newblock {\em At Data Nucl Data Tables}, 39(1):1--156, 1988.

\bibitem{Wahl2002}
A.~C. Wahl.
\newblock Systematics of fission-product yields.
\newblock {\em LA--13928}, page~64, 2002.

\bibitem{Katakura2003}
J.~Katakura.
\newblock {A systematics of fission product mass yields with 5 gaussian
  functions}.
\newblock Technical report, Japan At. Energy Res. Inst., Chiba, 2003.

\bibitem{Madland1982}
D.~G. Madland and J.~R. Nix.
\newblock New calculation of prompt fission neutron spectra and average prompt
  neutron multiplicities.
\newblock {\em Nucl Sci Eng}, 81(2):213--271, 1982.

\bibitem{Lemaire2005}
S.~Lemaire, P.~Talou, T.~Kawano, M.~B. Chadwick, and D.~G. Madland.
\newblock {Monte Carlo approach to sequential neutron emission from fission
  fragments}.
\newblock {\em Phys Rev C}, 72:024601, Aug 2005.

\bibitem{Talou2021}
P.~Talou, I.~Stetcu, P.~Jaffke, M.E. Rising, A.E. Lovell, and T.~Kawano.
\newblock Fission fragment decay simulations with the {CGMF} code.
\newblock {\em Comput Phys Commun}, 269:108087, 2021.

\bibitem{Litaize2010}
O.~Litaize and O.~Serot.
\newblock {Investigation of phenomenological models for the Monte Carlo
  simulation of the prompt fission neutron and $\ensuremath{\gamma}$ emission}.
\newblock {\em Phys Rev C}, 82:054616, Nov 2010.

\bibitem{Randrup2009}
J.~Randrup and R.~Vogt.
\newblock Calculation of fission observables through event-by-event simulation.
\newblock {\em Phys Rev C}, 80:024601, Aug 2009.

\bibitem{Vogt2009}
R.~Vogt, J.~Randrup, J.~Pruet, and W.~Younes.
\newblock Event-by-event study of prompt neutrons from
  $^{239}\mathrm{Pu}(n,f)$.
\newblock {\em Phys Rev C}, 80:044611, Oct 2009.

\bibitem{Schmidt2016}
K.-H. Schmidt, B.~Jurado, C.~Amouroux, and C.~Schmitt.
\newblock General description of fission observables: {GEF} model code.
\newblock {\em Nucl Data Sheets}, 131:107--221, 2016.
\newblock Special Issue on Nuclear Reaction Data.

\bibitem{Tudora2018}
A.~Tudora, F.-J. Hambsch, and V.~Tobosaru.
\newblock Revisiting the residual temperature distribution in prompt neutron
  emission in fission.
\newblock {\em Eur Phys J A}, 54(5):87, May 2018.

\bibitem{Okumura2018}
S.~Okumura, T.~Kawano, P.~Jaffke, P.~Talou, and S.~Chiba.
\newblock {$^{235}$U(n,f) Independent fission product yield and isomeric ratio
  calculated with the statistical Hauser-Feshbach theory}.
\newblock {\em J Nucl Sci Technol}, 55(9):1009--1023, 2018.

\bibitem{Tudora2006}
A.~Tudora.
\newblock Experimental prompt fission neutron ``sawtooth" data described by the
  ``point by point" model.
\newblock {\em Ann Nucl Energy}, 33(11):1030--1038, 2006.

\bibitem{Koning2012}
A.J. Koning and D.~Rochman.
\newblock Modern nuclear data evaluation with the {TALYS} code system.
\newblock {\em Nucl Data Sheets}, 113(12):2841--2934, 2012.
\newblock Special Issue on Nuclear Reaction Data.

\bibitem{Sadhukhan2016}
J.~Sadhukhan, W.~Nazarewicz, and N.~Schunck.
\newblock Microscopic modeling of mass and charge distributions in the
  spontaneous fission of $^{240}\mathrm{Pu}$.
\newblock {\em Phys Rev C}, 93:011304, Jan 2016.

\bibitem{Tanimura2017}
Y.~Tanimura, D.~Lacroix, and S.~Ayik.
\newblock Microscopic phase-space exploration modeling of $^{258}\mathrm{Fm}$
  spontaneous fission.
\newblock {\em Phys Rev Lett}, 118:152501, Apr 2017.

\bibitem{Lemaitre2019}
J.-F. Lema\^{\i}tre, S.~Goriely, S.~Hilaire, and J.-L. Sida.
\newblock Fully microscopic scission-point model to predict fission fragment
  observables.
\newblock {\em Phys Rev C}, 99:034612, Mar 2019.

\bibitem{Bulgac2019}
A.~Bulgac, S.~Jin, K.~J. Roche, N.~Schunck, and I.~Stetcu.
\newblock Fission dynamics of $^{240}\mathrm{Pu}$ from saddle to scission and
  beyond.
\newblock {\em Phys Rev C}, 100:034615, Sep 2019.

\bibitem{Bulgac2020}
A.~Bulgac, S.~Jin, and I.~Stetcu.
\newblock Nuclear fission dynamics: Past, present, needs, and future.
\newblock {\em Front Phys}, 8, 2020.

\bibitem{Zhao2023}
K.~Zhao, Y.J. He, Z.X. Li, L.L. Liu, C.W. Shen, Y.J. Chen, and X.Z. Wu.
\newblock The description of dynamical fission process using improved quantum
  molecular dynamics model incorporated with microscopic potential energy
  surface.
\newblock {\em Phys Lett B}, 839:137817, 2023.

\bibitem{Asano2004}
T.~Asano, T.~Wada, M.~Ohta, T.~Ichikawa, S.~Yamaji, and H.~Nakahara.
\newblock Dynamical calculation of multi-modal nuclear fission of fermium
  nuclei.
\newblock {\em J Nucl Radiochem Sci}, 5(1):1--5, 2004.

\bibitem{Randrup2011}
J.~Randrup and P.~M\"oller.
\newblock Brownian shape motion on five-dimensional potential-energy
  surfaces:nuclear fission-fragment mass distributions.
\newblock {\em Phys Rev Lett}, 106:132503, Mar 2011.

\bibitem{Aritomo2014}
Y.~Aritomo, S.~Chiba, and F.~Ivanyuk.
\newblock Fission dynamics at low excitation energy.
\newblock {\em Phys Rev C}, 90:054609, Nov 2014.

\bibitem{Pasca2016}
H.~Pasca, A.V. Andreev, G.G. Adamian, and N.V. Antonenko.
\newblock Possible origin of transition from symmetric to asymmetric fission.
\newblock {\em Phys Lett B}, 760:800--806, 2016.

\bibitem{Sierk2017}
A.~J. Sierk.
\newblock Langevin model of low-energy fission.
\newblock {\em Phys Rev C}, 96:034603, Sep 2017.

\bibitem{Ishizuka2017}
C.~Ishizuka, M.~D. Usang, F.~A. Ivanyuk, J.~A. Maruhn, K.~Nishio, and S.~Chiba.
\newblock {Four-dimensional Langevin approach to low-energy nuclear fission of
  $^{236}${U}}.
\newblock {\em Phys Rev C}, 96:064616, Dec 2017.

\bibitem{Jaffke2018}
P.~Jaffke, P.~M\"oller, P.~Talou, and A.~J. Sierk.
\newblock Hauser-feshbach fission fragment de-excitation with calculated
  macroscopic-microscopic mass yields.
\newblock {\em Phys Rev C}, 97:034608, Mar 2018.

\bibitem{Carjan2019}
N.~Carjan, F.~A. Ivanyuk, and Yu.~Ts. Oganessian.
\newblock Fission of superheavy nuclei: Fragment mass distributions and their
  dependence on excitation energy.
\newblock {\em Phys Rev C}, 99:064606, Jun 2019.

\bibitem{Mumpower2020}
M.~R. Mumpower, P.~Jaffke, M.~Verriere, and J.~Randrup.
\newblock Primary fission fragment mass yields across the chart of nuclides.
\newblock {\em Phys Rev C}, 101:054607, May 2020.

\bibitem{Usang2019}
M.~D. Usang, F.~A. Ivanyuk, C.~Ishizuka, and S.~Chiba.
\newblock {Correlated transitions in TKE and mass distributions of fission
  fragments described by 4-D Langevin equation}.
\newblock {\em Sci Rep}, 9(1):1525, Feb 2019.

\bibitem{Ishizuka2020}
C.~Ishizuka, X.~Zhang, M.~D. Usang, F.~A. Ivanyuk, and S.~Chiba.
\newblock Effect of the doubly magic shell closures in $^{132}\mathrm{Sn}$ and
  $^{208}\mathrm{Pb}$ on the mass distributions of fission fragments of
  superheavy nuclei.
\newblock {\em Phys Rev C}, 101:011601, Jan 2020.

\bibitem{Maruhn1972}
J.~Maruhn and W.~Greiner.
\newblock The asymmetric two center shell model.
\newblock {\em Z Physik}, 251(5):431--457, Oct 1972.

\bibitem{Ivanyuk2018}
F.~A. Ivanyuk, C.~Ishizuka, M.~D. Usang, and S.~Chiba.
\newblock Temperature dependence of shell corrections.
\newblock {\em Phys Rev C}, 97:054331, May 2018.

\bibitem{Pashkevich1971}
V.V. Pashkevich.
\newblock On the asymmetric deformation of fissioning nuclei.
\newblock {\em Nucl Phys A}, 169(2):275--293, 1971.

\bibitem{Kelson1964}
I.~Kelson.
\newblock Dynamic calculations of fission of an axially symmetric liquid drop.
\newblock {\em Phys Rev}, 136:B1667--B1673, Dec 1964.

\bibitem{Davies1976}
K.~T.~R. Davies, A.~J. Sierk, and J.~R. Nix.
\newblock Effect of viscosity on the dynamics of fission.
\newblock {\em Phys Rev C}, 13:2385--2403, Jun 1976.

\bibitem{Blocki1978}
J.~Blocki, Y.~Boneh, J.R. Nix, J.~Randrup, M.~Robel, A.~J. Sierk, and W.J.
  Swiatecki.
\newblock One-body dissipation and the super-viscidity of nuclei.
\newblock {\em Ann Phys}, 113(2):330--386, 1978.

\bibitem{Sierk1980}
A.~J. Sierk and J.~R. Nix.
\newblock Fission in a wall-and-window one-body-dissipation model.
\newblock {\em Phys Rev C}, 21:982--987, Mar 1980.

\bibitem{Adeev2005}
G.D. Adeev, Alexander Karpov, Pavel Nadtochy, and D.V. Vanin.
\newblock Multidimensional stochastic approach to fission dynamic of excited
  nuclei.
\newblock {\em Phys Part Nucl}, 36(4):378--426, 07 2005.

\bibitem{Okumura2020}
S.~Okumura, T.~Kawano, and S.~Chiba.
\newblock {The fission yield calculations with Langevin model, Hauser-Feshbach
  statistical decay, and beta decay}.
\newblock {\em EPJ Web Conf.}, 239:03005, 2020.

\bibitem{Brosa1990}
U.~Brosa, S.~Grossmann, and A.~Muller.
\newblock Nuclear scission.
\newblock {\em Physics Reports}, 197(4):167--262, 1990.

\bibitem{Schillebeeckx1992}
P.~Schillebeeckx, C.~Wagemans, A.J. Deruytter, and R.~Barthelemy.
\newblock {Comparative study of the fragments' mass and energy characteristics
  in the spontaneous fussion of $^{238}$Pu, $^{240}$Pu and $^{242}$Pu and in
  the thermal-neutron-induced fission of $^{239}$Pu}.
\newblock {\em Nucl Phys A}, 545(3):623--645, 1992.

\bibitem{Ohsawa1999}
T.~Ohsawa, T.~Horiguchi, and H.~Hayashi.
\newblock Multimodal analysis of prompt neutron spectra for $^{237}${Np}(n,f).
\newblock {\em Nucl Phys A}, 653(1):17--26, 1999.

\bibitem{Ohsawa2000}
T.~Ohsawa, T.~Horiguchi, and M.~Mitsuhashi.
\newblock {Multimodal analysis of prompt neutron spectra for $^{238}$Pu(sf),
  $^{240}$Pu(sf), $^{242}$Pu(sf) and $^{239}$Pu(n$_{\rm th}$,f)}.
\newblock {\em Nucl Phys A}, 665(1):3--12, 2000.

\bibitem{Okumura2022}
S.~Okumura, T.~Kawano, A.~E. Lovell, and T.~Yoshida.
\newblock {Energy dependent calculations of fission product, prompt, and
  delayed neutron yields for neutron induced fission on $^{235}$U, $^{238}$U,
  and $^{239}$Pu}.
\newblock {\em J Nucl Sci Technol}, 59(1):96--109, 2022.

\bibitem{Fujio2023}
K.~Fujio, A.~Al-Adili, F.~Nordstrom, J.-F. Lemaitre, S.~Okumura, S.~Chiba, and
  A.~Koning.
\newblock {TALYS calculations of prompt fission observables and independent
  fission product yields for the neutron-induced fission of $^{235}$U}.
\newblock {\em Eur Phys J A}, 59(8):178, Aug 2023.

\bibitem{Dematte1997}
L.~Dematte, C.~Wagemans, R.~Barthelemy, P.~D'hondt, and A.~Deruytter.
\newblock {Fragments' mass and energy characteristics in the spontaneous
  fission of $^{236}$Pu, $^{238}$Pu, $^{240}$Pu, $^{242}$Pu, and $^{244}$Pu}.
\newblock {\em Nucl Phys A}, 617(3):331--346, 1997.

\bibitem{Wagemans1988}
C.~Wagemans.
\newblock Contemporary fission.
\newblock {\em Proc. 5th Int. Symp. on Nucleon Induced Reactions}, 1988.

\bibitem{Wagemans1989}
C.~Wagemans.
\newblock On the necessity of alternative methods to determine sample
  thicknesses and masses.
\newblock {\em Nuclear Inst and Methods in Physics Research A}, 282(1):4--9,
  1989.

\bibitem{Surin1972}
V.~M. Surin, A.~I. Sergachev, N.~I. Rezchikov, and B.~D. Kuz'minov.
\newblock Yields and kinetic energies of fragments at the fission of
  $^{233}${U} and $^{239}${Pu} by 5.5 and 15 {MeV} neutrons.
\newblock {\em Yadern. Fiz. 14: No. 5, 935-8(Nov 1971).}, 1 1971.

\bibitem{Tsuchiya2000}
C.~Tsuchiya, Y.~Nakagome, H.~Yamana, H.~Moriyama, K.~Nishio, I.~Kanno, K.~Shin,
  and I.~Kimura.
\newblock Simultaneous measurement of prompt neutrons and fission fragments for
  $^{239}${Pu}(n$_{\rm th}$,f).
\newblock {\em J Nucl Sci Technol}, 37(11):941--948, 2000.

\bibitem{Shimada2021}
K.~Shimada, C.~Ishizuka, F.~A. Ivanyuk, and S.~Chiba.
\newblock Dependence of total kinetic energy of fission fragments on the
  excitation energy of fissioning systems.
\newblock {\em Phys Rev C}, 104:054609, Nov 2021.

\bibitem{Mueller1984}
R.~M\"uller, A.~A. Naqvi, F.~K\"appeler, and F.~Dickmann.
\newblock Fragment velocities, energies, and masses from fast neutron induced
  fission of $^{235}\mathrm{U}$.
\newblock {\em Phys Rev C}, 29:885--905, Mar 1984.

\bibitem{Naqvi1986}
A.~A. Naqvi, F.~K\"appeler, F.~Dickmann, and R.~M\"uller.
\newblock Fission fragment properties in fast-neutron-induced fission of
  $^{237}\mathrm{Np}$.
\newblock {\em Phys Rev C}, 34:218--225, Jul 1986.

\bibitem{Tudora2015}
A.~Tudora, F.-J. Hambsch, I.~Visan, and G.~Giubega.
\newblock {Comparing different energy partitions at scission used in prompt
  emission model codes GEF and Point-by-Point}.
\newblock {\em Nucl Phys A}, 940:242--263, 2015.

\bibitem{Thulliez2019}
L.~Thulliez, O.~Litaize, O.~Serot, and A.~Chebboubi.
\newblock Neutron and $\ensuremath{\gamma}$ multiplicities as a function of
  incident neutron energy for the $^{237}\mathrm{Np}$(n,f) reaction.
\newblock {\em Phys Rev C}, 100:044616, Oct 2019.

\bibitem{Kawano2021}
T.~Kawano, S.~Okumura, A.~E. Lovell, I.~Stetcu, and P.~Talou.
\newblock Influence of nonstatistical properties in nuclear structure on
  emission of prompt fission neutrons.
\newblock {\em Phys Rev C}, 104:014611, Jul 2021.

\end{thebibliography}
